\documentclass{pnastwo}

\usepackage[dvips]{graphicx}
\usepackage{color}

\usepackage{amssymb}
\usepackage{amsfonts}
\usepackage{amsmath}
\usepackage[linesnumbered,ruled,vlined]{algorithm2e}
\usepackage{color}

\usepackage{caption}
\usepackage{subcaption}

\begin{document}
	\title{Mind Control as a Guide for the Mind}
	
	
	
	
	\author{John D. Medaglia\affil{1} {Department of Psychology, University of Pennsylvania, Philadelphia, PA, 19104 USA},
		Perry Zurn \affil{2}{Center for Curiousity, University of Pennsylvania, Philadelphia, PA, 19104 USA}\affil{3}{Department of Philosophy, American University, Washington, DC, 20016 USA}, Walter Sinnott-Armstrong \affil{4}{Department of Philosophy and Kenan Institute for Ethics, Duke University, Durham, NC, 27708 USA},
		Danielle S. Bassett \affil{5}{Department of Bioengineering, University of Pennsylvania, Philadelphia, PA, 19104 USA}\affil{6}{Department of Electrical \& Systems Engineering, University of Pennsylvania, Philadelphia, PA, 19104 USA}
	}
	\contributor{}
	
	\maketitle
	
	\begin{article}

\begin{abstract}
	The human brain is a complex network that supports mental function. The nascent field of network neuroscience applies tools from mathematics to neuroimaging data in the hopes of shedding light on cognitive function. A critical question arising from these empirical studies is how to modulate a human brain network to treat cognitive deficits or enhance mental abilities. While historically a number of tools have been employed to modulate mental states (such as cognitive behavioral therapy and brain stimulation), theoretical frameworks to guide these interventions -- and to optimize them for clinical use -- are fundamentally lacking.  One promising and as-yet under-explored approach lies in a sub-discipline of engineering known as \emph{network control theory}. Here, we posit that network control fundamentally relates to mind control, and that this relationship highlights important areas for future empirical research and opportunities to translate knowledge into practical domains. We clarify the conceptual intersection between neuroanatomy, cognition, and control engineering in the context of network neuroscience. Finally, we discuss the challenges, ethics, and promises of mind control.
\end{abstract}

		\keywords{brain network | controllability | network science | diffusion tractography | cognitive control}

\section*{Introduction}

Mind control is a common plot device in many genres of fiction. Its ubiquity is perhaps unsurprising: the prospect of the explicit, full control of the mind evokes alluring and startling possibilities. Fictional mind control often takes implausible forms: telepathy, magical interventions, and nefarious schemes of authoritarian organizations. In more biologically-inspired plot lines, mind control is delivered by devices implanted in the subject's brain, as depicted in \textit{The Matrix}: these devices manipulate neurophysiological processes resulting in a change of mental state. 

In reality, mind control encompasses numerous means for influencing the mind. This includes effects mediated through the senses. Sense-mediated effects can include positive influences, such as updating one's beliefs based on presented evidence, or ``nudging'' someone to make healthy decisions via environmental manipulation \cite{thaler2014choice}. They can also be more insidious, as in the case of propaganda or brain washing. Beyond social and environmental means, mind control can also result from direct neural stimulation. Neural stimulation can include subtle modulation via pharmacological agents or more direct manipulations with brain stimulation that result in neural discharges. The last few decades have seen a steady increase in the use of implanted devices to assist individuals with major mental disorders. The ubiquitous nature of social forms of control and increasing prevalence of neural devices motivates important questions about the control of brain processes and, by extension, mental functions.

Developing a true science of mind control could benefit from the engineering discipline known as ``control theory'', which addresses the question of how to guide complex systems from one state to another \cite{tk:80}. As a commonplace example, control systems in a modern airliner ensure that the aircraft stays aloft by automatically adjusting the plane's pitch, roll, and yaw to compensate for the turbulence in the air \cite{liu2011controllability}. Like a plane, the brain is a physical system that is characterized by specific states: in this case, patterns of neural activity. The control or guidance of the brain from one state to another can either be intrinsic (the brain controls itself \cite{gu2015controllability}) or extrinsic (the brain is guided externally, for example by brain stimulation \cite{muldoon2016stimulation}). 

Beyond these naturalistic forms of mind control, stimulation-based interventions have been developed for clinical cohorts, offering a powerful link between engineering and neuroscience. The use of control theoretic approaches are particularly prevalent in the booming field of neuroprosthetics \cite{humayun2013recent,arts2012review}, which can be used to treat motor disorders such as Parkinson's disease \cite{Schiff2012}. Here, control theory dictates efficient strategies for deep brain stimulation to regulate motor functions mediated by subcortical areas. Similar efforts have been developed to treat obsessive compulsive disorder \cite{figee2013deep} and depression \cite{mayberg2005deep}, suggesting their utility across functional domains. Moreover, neural control is administered through noninvasive yet powerful techniques such as transcranial magnetic stimulation \cite{hallett2000transcranial}.

To better describe how control theory from engineering can inform our understanding of mind control, it is useful to invoke a network approach \cite{gu2015controllability}. In a network perspective, neural components (such as individual neurons or entire brain regions) are treated as \emph{nodes} and connections between these components are treated as network \emph{edges} \cite{Bassett2006,Bullmore2009,Bullmore2011}. By using this conceptual framework, as well as the mathematical formalism that accompanies it, we can explicitly study how control input applied to one brain component can impact the rest of the network \cite{gu2015controllability,betzel2016optimally}. Indeed, the network formalism allows us to capitalize on recent advances in network control \cite{fp-sz-fb:13q}, which may prove useful in developing principled strategies that affect the mind. In particular, these strategies can theoretically guide neural systems, and thus cognition, toward target states.

In this paper, we aim to briefly review recently developed concepts and bodies of literature pertinent to a formal science of mind control. First, we introduce the general notion of control in the engineer's sense. Then, we discuss recent extensions of this notion to network control. We highlight what a control theoretic approach to the brain and mind implies more broadly. We close with a discussion of social and ethical considerations that will emerge as mind control develops. While we leave mathematical details to other excellent texts, we seek throughout to provide the reader with intuitions that are sufficient to consider the state of the art, promises, and challenges that lie ahead at the frontiers of mind control.

\section*{Control for networked systems}

In the broadest sense, ``control'' is any form of physical influence from one entity to another; in an engineering sense, control indicates using input to move a system from an initial state to a target state. Typically, control is exerted by a controller $C$ via control input, $u$, to change the state of a system (or ``plant" $P$ in engineering) (Fig. 1). The control strategy may minimize the distance between the observed state of the system and the target state, often while also maximizing technical simplicity and minimizing input \cite{lee1967foundations}. For example, the plant could be an aircraft, the controller could be the aircraft's turbine engines, and the control input induces thrust. The aircraft has sensors that monitor its state and continuously report that information to the controller, which in turn adapts its strategy to minimize the distance between the aircraft's current trajectory and the target trajectory.

\begin{figure} [h]
	\centerline{\includegraphics[width=3.5in]{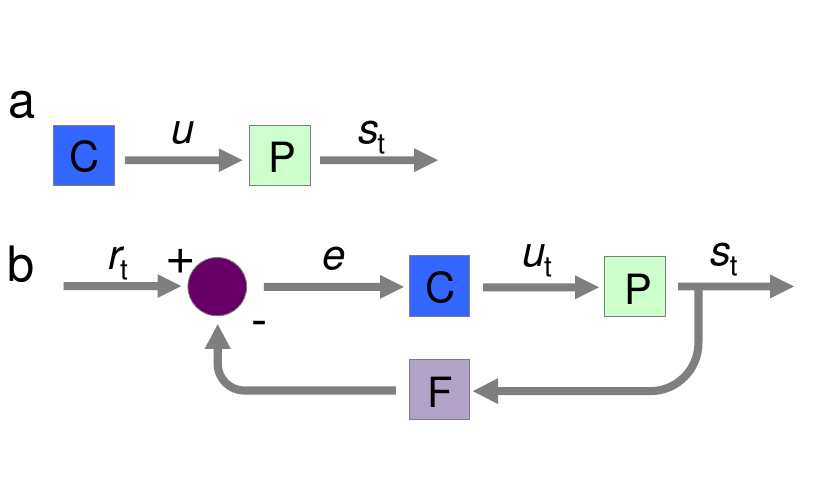}}
	\caption{\textbf{Control Theory}. Generic notions of control. (a): A classic open-loop control scheme. Traditionally, a controller (e.g., a system in the environment or designed device) delivers an input $u$ to the system under control (usually termed the ``plant'' $P$) to influence the system state $s_{t}$. (b): A classic closed-loop feedback control scheme. The goal is to guide the system ($P$) to a reference value $r_{t}$. The system state $s_{t}$ is fed back through a sensor measurement $F$ to compare to the reference value $r_{t}$. The controller $C$ then takes the error $e$ (difference) between the reference and the output to change the control inputs $u$ to the system under control ($P$). }\label{genericcontrol}
\end{figure}

A challenge in control theory is to design strategies that interact naturally with a system's structure and dynamics. A simple example is a child on a swing. In this system, an input (a push) should be applied from the apex of the swing. A small push at the apex is more effective in accelerating the child than a large push at the bottom of the swing. This fact illustrates an important principle for control: we should seek solutions that involve low physical costs with relatively high accuracy and precision in achieving our control goals \cite{tk:80}. In more complex networked systems, we must understand the structure and dynamics of the system as precisely as possible to identify effective control strategies.

In the context of human cognition, the plant (system) is the neural tissue that supports cognition -- e.g., a single neuron, an ensemble of neurons, a collection of brain regions, or the whole brain \cite{stigen2011controlling,Schiff2012,gu2015controllability,nabi2011single}.  In deep brain stimulation in Parkinson's disease, the controller is the stimulation device, the input is an applied current, and the system is a portion of the basal ganglia (often the globus pallidus or putamen). The neural reference state is often represented in the frequency domain, and the neural control goal is to achieve a target state of basal ganglia activity -- and the rest of the motor system with which it interacts -- that facilitates unimpaired motor movements \cite{Schiff2012,sarma2010using}. Beyond simple motor function, neural control can be used to treat obsessive compulsive disorder, where patients engage in repetitive behaviors following overwhelming urges to do so. Deep brain stimulation to subcortical circuitry effectively reduces symptoms of the disorder and substantially improves quality of life \cite{greenberg2006three}, putatively by normalizing neural activity in a fronto-striatal circuit \cite{figee2013deep}.

One essential advantage in applying control theory to natural systems is that real stimuli tend to affect many parts of systems, not just those directly targeted. Indeed, one challenge in neural stimulation is that input or stimulation to one area affects others \cite{muldoon2016stimulation}. Progress in understanding this complexity requires theory, mathematics, and finely resolved network data to represent the interconnections between brain areas that propagate input \cite{gu2017optimal}. Control theory provides us with principled approaches to identify strategies that influence states in the context of indirect interactions within the system \cite{kailath1980linear}. Thus, in conjunction with brain \emph{connectomics} \cite{sporns2005human}, control theory is particularly promising in our efforts to guide the mind. Indeed, as we will discuss, control theory may offer a much-needed framework for the use of many neural stimulation techniques to influence cognitive function in health \cite{miniussi2013modelling} and disease \cite{brunoni2012clinical,teneback2015changes,nitsche2016tdcs}.

The study of how to control a complex network is commonly referred to as \emph{network control theory} \cite{liu2011controllability,Ruths2014}. A networked system is represented as a ``graph'' of interconnected elements, where nodes represent components of a system and edges represent connections or interactions between nodes. In the brain, nodes typically represent neural elements (e.g., neurons or brain regions), and edges represent connections between nodes (e.g., axons, information, or white matter tracts \cite{medaglia2015cognitive}). Network control theory is the study of how to design control inputs to a network that can be used to guide the system from an initial state to a target state (see Fig. 2) \cite{betzel2016optimally,gu2017optimal}.

Network control is conceptually appropriate for studying how to affect the mind using time-varying input in specific parts of the brain, thereby inducing trajectories of neural dynamics via anatomical connections to achieve precise goals at a low cost. Here, the goal is to induce changes in neural states over time that support cognitive functions. This highlights an important dual nature to mind control: the \emph{neural} control goal is to induce movement from one neural state to another. The \emph{psychological} control goal presumably depends on neural states, but can be sensibly discussed in its own terms. Indeed, if no psychological, physical, or social consequences were associated with depression, there would be no such thing as a mind control goal for depression. 

\begin{figure} [h]
	\centerline{\includegraphics[width=3.5in]{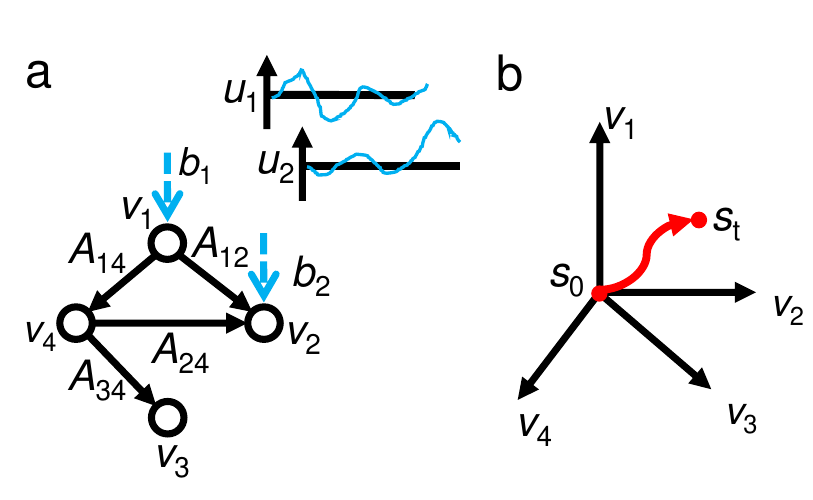}}
	\caption{\textbf{Network control.} (a): Network nodes $v$ are connected through edges $A_{ij}$. Control input $u$ can be administered to nodes via controllers $b$ in time-varying patterns to obtain the best control strategy given knowledge of the system and the nature of the controller.  (b) Control input can be designed to guide a network from an initial state $(s_{0})$ to a target state later in time $(s_{t})$. Figure adapted with permission from \cite{liu2011controllability}. }\label{networkcontrol}
\end{figure}

\section*{Brain controllability and guiding the mind}

In network control theory applied to the brain, we can consider any case involving structural pathways in the brain (e.g., individual axons or white matter bundles) that connect neural elements (e.g., single neurons or groups of neurons) to one another. Initial applications of network control theory to the human connectome suggest that the brain may employ distinct control strategies to guide mental processes \cite{gu2015controllability}. Three well-known cognitive systems display different patterns of structural connectivity estimated by white matter tractography, and those patterns facilitate different types of control. For example, the fronto-parietal system is a set of regions known to facilitate the human's ability to switch between different tasks \cite{miller2001prefrontal,Dosenbach2007}. Interestingly, this system exhibits relatively sparse anatomical connectivity with the rest of the brain. Results from network control theory suggest that the fronto-parietal system is optimized for moving the brain into difficult-to-reach states \cite{gu2015controllability} along an ``energy landscape'', which defines the possible states and transitions of the network \cite{watanabe2014energy}. Dorsal and ventral attention systems \cite{Corbetta2002,Sestieri2014} have neither dense nor sparse connectivity and may be optimized for integrating or segregating other parts of the brain. Finally, the brain's baseline system -- the so-called ``default mode'' \cite{raichle2001default,fox2006spontaneous,Fox2007} -- is strongly structurally connected with the brain and may drive the brain to many easy-to-reach states. These findings suggest that the brain is organized into anatomically distinct control systems \cite{gu2015controllability}. 

Based on this previous work, it is plausible that distinct brain regions could be candidate targets for interventions that influence brain dynamics into distinct dynamic trajectories. Supporting this notion, evidence from numerical simulations suggests that the structural connections emanating from a brain region directly impact the transmission of stimulation from the targeted area to the rest of the brain \cite{muldoon2016stimulation}. In that study, stimulation to regions in the default mode imparted large global change in brain activity, suggesting the importance of considering individual differences in white matter tracts in brain stimulation protocols. In addition to being a potential target for brain stimulation, regions in the default mode may also play a role in homeostasis following stimulation, as evidence suggests that they are the least energetically costly target state \cite{betzel2016optimally}.

While these initial studies have focused on the large-scale connectome of the human, network control theory can be applied more broadly. Presumably, cognition -- and arguably its control -- occurs over multiple spatial resolutions \cite{bassett2011understanding}, offering distinct targets for network control in support of mental function (see Fig. 3). Neuroimaging continues to clarify the macro scale organization of networks that supports cognitive processes \cite{medaglia2015cognitive}. However, to gain increasing control over cognitive function, a more fundamental characterization of cognition will be required. Specifically, in emerging theoretical work on \emph{neural} control across the structural connectome, the mapping between brain dynamics and specific cognitive processes will be critical to inform \emph{psychological} (i.e., ``mind") control. At the level of organization associated with many cognitive functions, neural control goals can focus on affecting single neurons \cite{panzeri2010sensory} to -- in all likelihood, subtly -- influence cognition. In addition, control goals could involve neuronal ensembles \cite{sutskever2014sequence,merchant2013neural,stolier2016neural}, and large-scale distributed neural circuits \cite{Bassett2015,calhoun2014chronnectome,medaglia2015cognitive}. 

\newpage
\begin{figure*}[h]
	\centerline{\includegraphics[width=7in]{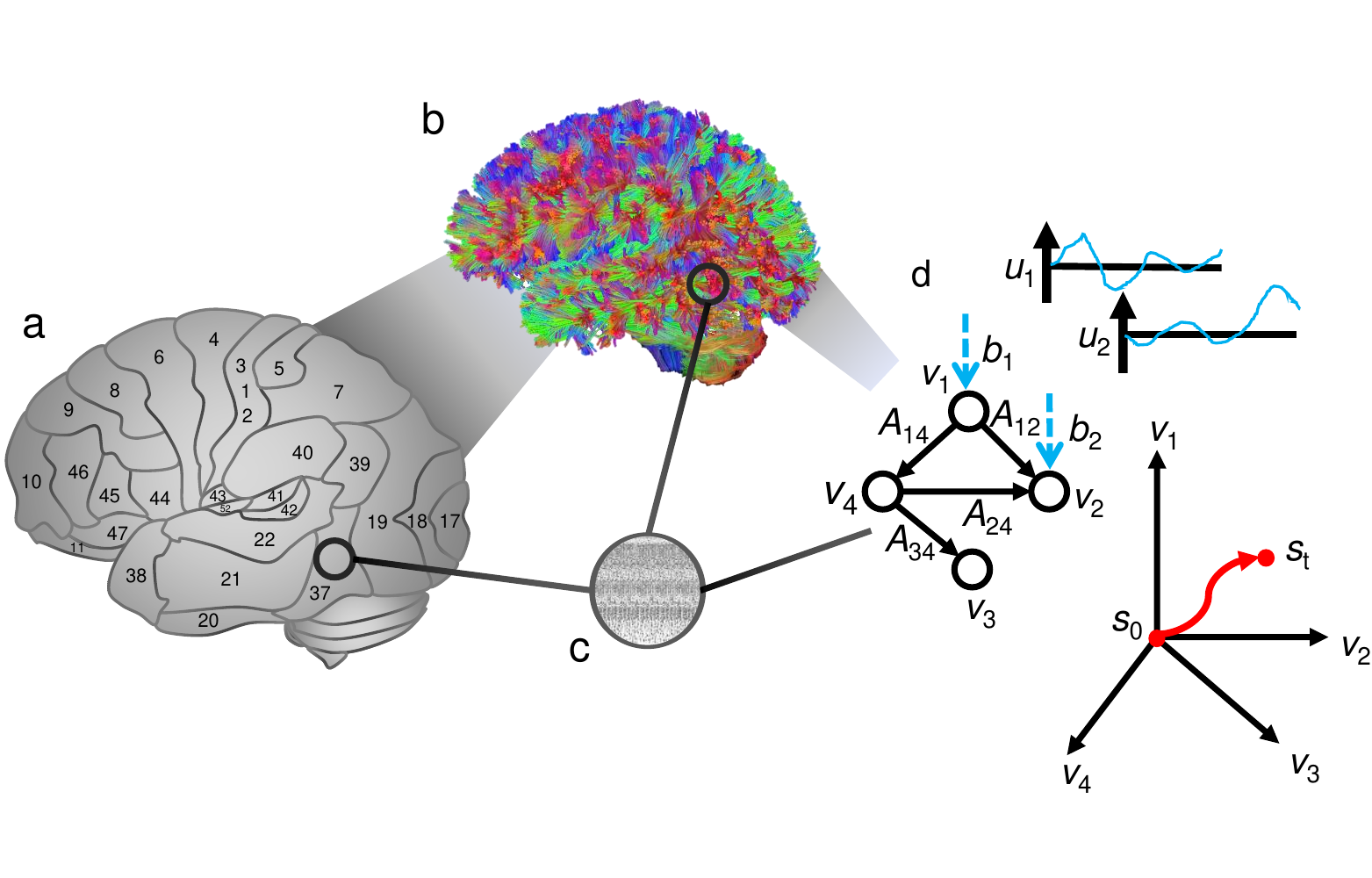}}
	\caption{\textbf{Brain control.} (a): The gross anatomical organization of the brain can be described by cytoarchitectonic regimes that serve distinct roles in neural computation \cite{Brodmann1909}. (b): Techniques such as diffusion weighted imaging can provide information about the macro-scale connectivity among brain regions (the ``connectome'' \cite{sporns2005human}). (c): Low level cellular organization facilitates information processing and is embedded within the macro-scale connectome. (d): The structural and dynamic organization of the brain at multiple scales can be represented as a networked system that can be guided using control input targeted to specific brain regions.}\label{braincontrol}
\end{figure*}
\newpage

\section*{Controlling specific mental functions}

\begin{figure*}[h!]
	\centerline{\includegraphics[width=7in]{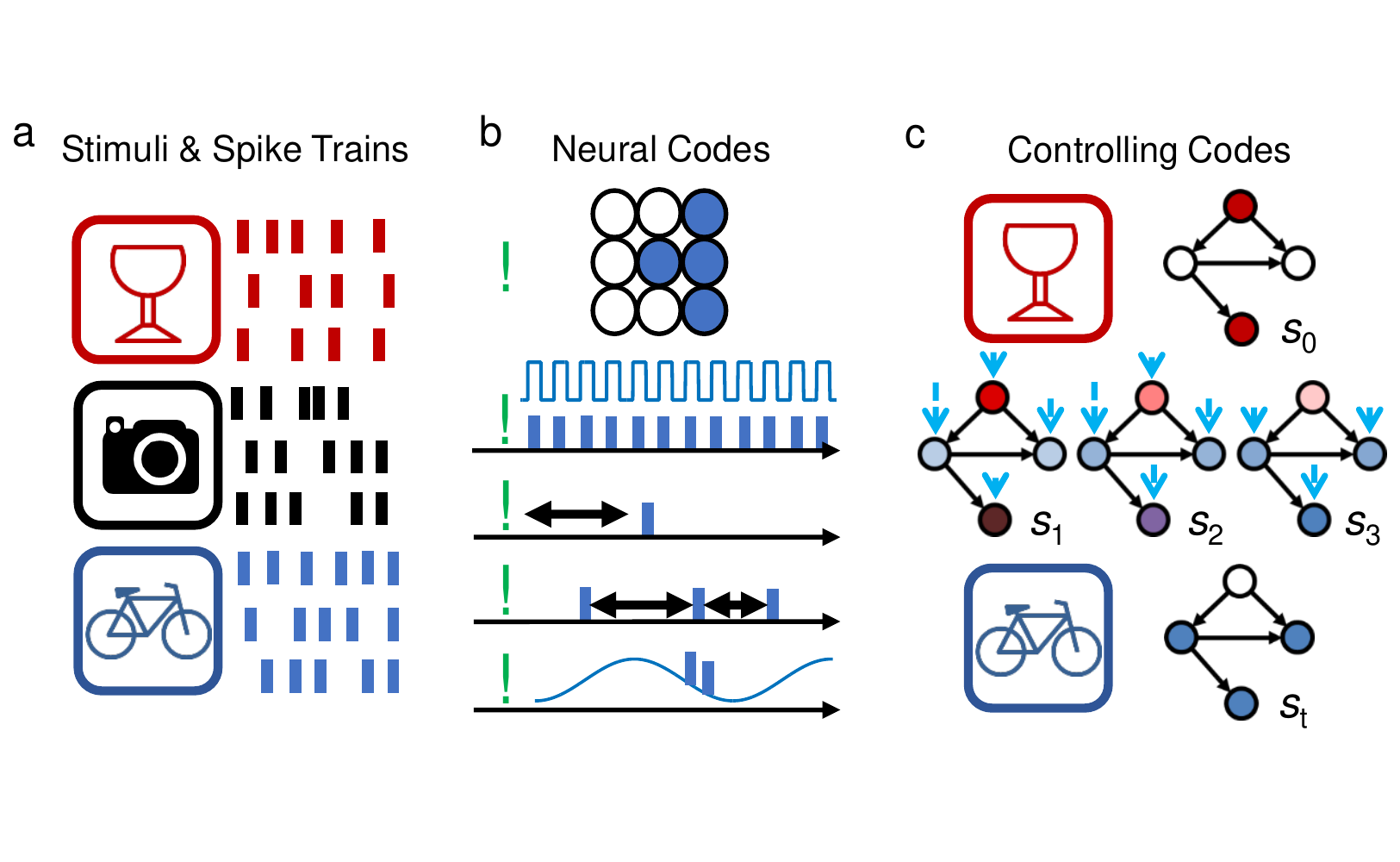}}
	\caption{\textbf{Neural codes and cognition.} A speculative schematic for the control of object perception. (a): Neurons transmit information in the form of neural discharges or ``spikes'' associated with stimuli. Here, neural spike trains in distinct colors represent different stimuli. (b): Neural spike trains can be analyzed to determine the nature of coding that supports distinct representations and processes that constitute stimuli. The green exclamation point represents an arbitrary stimulus that can (in principle) be represented by a number of possible codes. Top to bottom: a stimulus representation is maintained by a population of neurons, which may use frequency or ``rate'' coding, latency coding, interspike interval codes, or the phase of firing. (c): Top: the wine glass is represented by the initial neural state $s_{0}$ of two neurons in the four-neuron system. Middle to bottom: control input is applied to different neurons in varying quantities over time to induce a transition to a state representing the bicycle. The variable $s_{t}$ represents the state of the neural network at time $t$. Realizing the control strategy requires the right neural population, code, and manipulating apparatus. While we select a visual perception example for clarity, a similar intuition can apply to any neurocognitive process that involves temporal codes in populations of neurons.} \label{neuralcodes}
\end{figure*}

In reality, controlling mental functions would require a marriage between network control theory and the neural processes that enable cognition: namely, neural codes. Neural codes occur in several forms. These are typically thought to involve \emph{temporal} characteristics of neural firing within a certain \emph{population} of neurons \cite{shamir2014emerging}. The population may include relatively few neurons to increase the information processing capacity of a neural system \cite{zaslaver2015hierarchical}. Temporal rate codes can be represented in the frequency domain, where information is represented in how quickly neurons fire \cite{lisman2013theta,miura2012odor}. Temporal codes based on spike timing encode information in the delay between a stimulus and neural discharge \cite{ainsworth2012rates}. Temporal codes can be multiplexed to increase encoding capacity, disambiguate stimuli, and stabilize representations \cite{panzeri2010sensory} (see Fig.~\ref{neuralcodes} for a schematic example in visual perception).

\newpage

Defining control goals for neural applications is challenging because the mapping from neural states and state transitions to cognition is presently incomplete. There is tremendous spatiotemporal detail we could attempt to specify within neural systems. In addition, it is not known whether and to what extent the neural states underlying any particular cognitive process are memoryless \cite{csiszar2011information} or are achieved by passing through other states. Thus, a major challenge for any mind control goal is to identify what is important and what can be ignored. In clinical contexts, we often know that we want to guide behavior away from a particular pattern. Even if we can identify the neural state we want to avoid, however, we may not necessarily know the neural state we want to achieve \cite{Schiff2012}. Moreover, different individuals may use different cognitive strategies to approach the same tasks \cite{pressley2006cognitive}, suggesting that optimizing to individuals will require knowledge of neural representations within a single individual. Indeed, we must carefully evaluate our models of neural health, neuropathology, and neural repair with respect to our goals and the success of our interventions.

While the nature and relevance of cognitive ``representations'' remains an open issue \cite{van1995might}, a well-formed control strategy includes a clearly defined neural control goal and a mapping from neural state to cognitive representations and processes. This fact highlights a potentially important role for control theoretic approaches to drive fundamental discovery in cognitive neuroscience. To move cognitive neuroscience from correlation to causative models, a combination of direct manipulation and observation is required. Correlative neuroscience can help us identify specific neural patterns that may be associated with specific functions. Then, the possible causal nature of these relationships can be tested and validated with experimental perturbations using stimulation. If we tune control inputs to elicit desired cognitive effects, we can rule out some control strategies, allowing us to identify cognition-relevant ranges of neural responses. Once these ranges are identified, we can aim to understand the neural dynamics responsible for specific cognitive changes, then iteratively attempt better control strategies.

Another overarching challenge is the fact that the brain's anatomy and neural activity change over the lifespan \cite{sowell2003mapping}, during learning \cite{draganski2006temporal}, and in response to exogenous stimulation \cite{lan2016transcranial}. Thus, mind control goals are enacted in dynamic multiplexed neural networks \cite{betzel2016multi}, meaning that we should monitor not only changes in neural state due to stimulation, but also changes in anatomy and the basic neural organization of cognition  in our efforts to build robust control strategies for cognitive function. 

Furthermore, to enact practical mind control strategies, the time scale of effects is an important consideration. If a control strategy only affects cognition for moments, it may have no practical utility. Encouragingly, control strategies can be designed to administer continuous stimulation to treat motor \cite{Schiff2012} and cognitive symptoms \cite{karamintziou2015design} over months and years. Crucially, it may not always be necessary to administer continuous control to achieve long-lasting cognitive effects \cite{reis2009noninvasive,snowball2013long}. Thus, the use of control should be evaluated against the difficulty and cost of the design and administration of control in the face of alternative strategies. Some cognitive problems may require a few open-loop sessions to achieve durable effects, whereas others might require continuous and fast-timescale stimulation. Speculatively, this may depend on different degrees and types of neuroplasticity in different cognitive systems. Whether we can further optimize long-lasting effects with interventions based on network control theory remains to be seen.

While conjectural at this point in scientific history, it is possible that one could combine an understanding of neural codes, experimental manipulation via brain stimulation, and network control theory to design control strategies to achieve certain mental states. This defines the frontier of mind control.

\section*{Technologies for control}

Some opportunities for mind control exist in nascent stages of development that could benefit from analysis within network control theory. In clinical contexts, network control theory could inform cognitive repair strategies. It could also inform cognitive enhancement. At present, a number of technologies for neural stimulation are routinely used in experimental and clinical neuroscience (Fig. 4). Stimulation techniques typically capitalize on different spatiotemporal properties of electromagnetism. Microstimulation can influence the activity of single neurons. Most techniques used in cognitive neuroscience and clinics operate at a scale much coarser than a single neuron. For example, transcranial magnetic stimulation non-invasively induces current in cortical tissue, where it causes action potentials in large populations of neurons. With much less spatial precision, approaches such as transcranial direct current stimulation use current administered to the scalp that modulates neural firing thresholds in the cortex. Invasive approaches such as deep brain stimulation use implanted electrodes to influence local field potentials in groups of neurons. Techniques such as optogenetics allow us to deliver spatiotemporally precise control sequences  in tissues engineered to express light-sensitive ion channels \cite{deisseroth2011optogenetics}. The dynamics supporting cognitive function can be influenced by these techniques at a low energy and risk cost \cite{barker1985non}.

\newpage
\begin{figure} 
	\centerline{\includegraphics[width=3.5in]{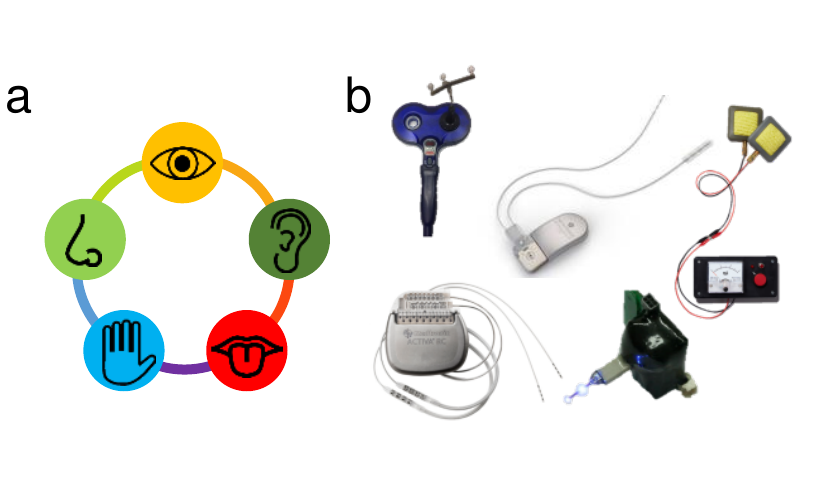}}
	\caption{\textbf{Forms of control.} (a): While not the primary emphasis of the current discussion, we note that one form of neural control is constantly mediated through sensation, which influences immediate experience and learning, and the basis of psychological treatments such as cognitive behavioral therapy. (b): Neural control can be administered via direct noninvasive or invasive neural stimulation devices. Clockwise from top left: MagStim transcranial magnetic stimulation coil, NeuroPace implanted stimulator, transcranial direct current stimulator, Triangle Biosystems optogenetic stimulator,  and Activa deep brain stimulator.}\label{methods}
\end{figure}
\newpage

Most applications of noninvasive electrical and magnetic stimulation are a form of open-loop control without feedback about internal neural states. This stands in contrast to closed-loop control, which involves feedback about internal neural states \cite{Schiff2012}. In open-loop control, control strategies are evaluated based on their success in influencing specific cognitive and behavioral functions \cite{silvanto2008state} or their efficacy in influencing clinical outcomes \cite{fox2014resting}. Here, the goal or ``reference" state is the cognitive status of the subject rather than the internal neural states. Studies in this area have revealed putative localizations for specific brain functions and led to FDA-approved therapies for depression, which involves significant emotional and cognitive disturbances \cite{george2010noninvasive}. Open-loop control research continues to provide substantial scientific and clinical benefits in behavioral neuroscience. It also illustrates that a \emph{psychological} control goal can be approached even when knowledge of specific neural codes is limited or completely absent. In open-loop control, it may be critical to calibrate methods for patients based on individualized neural patterns \cite{Finn2015}, using advanced measurements available at the hospital to optimize its use by the patient outside the hospital.

In closed-loop control, a critical component in the controller-observer (feedback sensor) cycle is the ability to detect states of the system. This is one reason that closed-loop control is easier to achieve invasively: we can better observe and control neural states when the sensors and controllers have direct access to neurons \cite{Schiff2012}. Techniques such as deep brain stimulation benefit from real-time detection of neural states, where the control strategy can adapt its input sequence based on the current states of the system \cite{Schiff2012}. At this scale, implanted microsensors record the local activity in neurons \cite{kozai2012ultrasmall}. Optogenetics can be paired with sensors to administer closed-loop control in translational models \cite{grosenick2015closed}. Although invasive strategies can generally better achieve closed-loop control, at higher spatiotemporal scales, spatially coarse states can be measured by techniques such as EEG \cite{mullen2015real}, and temporally coarse states can be measured with real-time fMRI \cite{zotev2014self}. Technologies including mobile EEG and functional near-infrared spectroscopy (fNIRS) can be useful in applied and rehabilitative settings where more cumbersome measurement tools are of limited use \cite{sitaram2016closed}. In biofeedback paradigms, the participant can observe his or her own measured neural states, and can adaptively learn to control these signals \cite{florin2014targeted}.

\section*{Good enough control}
To truly understand the means to control the mind suggests that one has sufficiently observed it \cite{kalman1960new, Schiff2012}. This is no simple task. Brain stimulation strategies for mind control are limited by knowledge of the underlying brain network architecture, the nature of neural codes, and the limitations of technology and computing \cite{Schiff2012}. All measurement techniques have spatiotemporal and other technological limitations \cite{sitaram2016closed}. Given the difficulties in measuring neural states, modeling system dynamics, and estimating signals within noise, simplifying assumptions can be useful to facilitate insights into intrinsic \cite{gu2015controllability} and extrinsic \cite{muldoon2016stimulation} control of brain function. Even with simplifying assumptions, studies suggest that the brain is very difficult, if not impossible, to control \cite{gu2015controllability,tarokh1992measures}. Thus, one can likely only ``guide'' the brain rather than ``control'' it in its entirety. 

Encouragingly, neural control strategies can be designed with limited information about neural systems \cite{Schiff2012}. Various techniques in neuroscience provide information about neural states with varying degrees of spatiotemporal precision. To date, efforts in mapping cognitive functions within the brain have relied on techniques ranging from invasive microstimulation and recording on limited sets of neurons to non-invasive imaging and computational techniques such as multi-voxel pattern analysis \cite{norman2006beyond}. In addition, transitions between states at a coarse scale can be studied using dynamic network approaches \cite{calhoun2014chronnectome,Kopell2014,medaglia2015cognitive}. Thus, while an ideal control strategy would include precise mapping between neural codes and the mind, numerous findings in brain stimulation research across neural scales suggest that the mind can be influenced in ways that provide interpretive and practical value. 

In particular, guiding the mind is perhaps most pragmatic in clinical scenarios where internal neural states are not directly measured, sensor feedback is difficult to maintain, and costs for comprehensive evaluation are prohibitive. For these scenarios, engineers are developing tools to influence the dynamic trajectory of the system using control strategies with limited access to the system and finite control input \cite{Motter2015}. More specifically, these techniques offer guidelines to steer the network into the intended state via cost-efficient strategies that simultaneously influence the system and allow the system's natural dynamics to help drive the system to the control goal \cite{tang2017control}. In the human brain, we could use models of neural activity \cite{sotero2007realistically,brown2004phase,kilpatrick2015wilson} to identify strategies that guide the brain to desirable states at a low cost by assessing which inputs will drive the brain's natural dynamics into an energy efficient trajectory. One notable issue is that some structures such as subcortical systems are difficult to access non-invasively, so we can seek to develop indirect control strategies using noninvasive stimulation \cite{denslow2005cortical} and knowledge of cortico-subcortical connectivity \cite{groppa2013subcortical}. Future studies utilize these and similar approaches to identify candidate strategies to guide brain network dynamics in the context of missing information and suboptimal control.

To close this speculative discussion, we mention a few additional limitations. First, we motivate these ideas based on the notion of ``structural controllability'': the control input is mediated through the nodes (brain regions) in the network, and its influences are relayed through edges (white matter tracts) to other nodes. In principle, other control strategies can be designed \cite{nowzari2016analysis}, including (i) modifying network edges by introducing bypasses for damaged tissue via neuroprosthetics \cite{bouton2016restoring} and (ii) modifying local regional dynamics using pharmacology \cite{zhao2012systems}. Second, we note that mind control holds no special ontological status relative to motor control. Indeed, the distinction between motor and cognitive processes may simply be an artifact of historical emphasis \cite{rosenbaum2005cinderella}. Accordingly, trade-offs between motor and cognitive function are evident in Parkinson's disease when a single site is stimulated and elicits improvements in motor performance at the cost of increased impulsivity \cite{frank2007hold}. If both motor and cognitive function are fundamentally computational processes \cite{churchland1989neural}, both can be informed by neural control engineering. However, there may be differences in the quality and degree of control required across these domains, as well as different stakes.

\section*{The ethics of brain control}
As efforts to guide complex brain processes advance, we will not only need new theoretical and technical tools. We will also face new societal, legal, and ethical challenges. Our best chance of meeting those challenges is through ongoing, rigorous discussion between scientists, ethicists, and policy makers.

Even at this early stage of the science of mind control, present and future ethical issues are important to consider. These issues pertain to developing research and clinical applications. In both cases, it is crucial to maximize benefits to society and protect against harm. In addition, ethical restrictions that apply in the use of mind control to treat dysfunction and alleviate suffering may differ from those that apply where mind control may be used to enhance typical function \cite{hamilton2011rethinking}.

Insofar as mind control has and will be undertaken in experimental and clinical contexts, four basic principles of medical and research ethics apply: nonmaleficence, beneficence, justice, and autonomy \cite{shamoo2009responsible}. Adhering to these principles is a first step toward ensuring that efforts to guide the mind enhance human welfare without violating human rights. Here we will take direct neural and magnetic stimulation as our primary points of emphasis. In addition, we will identify basic ethical issues that may be important to the more complex case of sensation-mediated stimulation, which includes the broader social and environmental forms of control that individuals experience in daily living.

\paragraph{Non-maleficence.}
The principle of nonmaleficence should supersede any implementation of control in experimental or clinical contexts. In developing and using these techniques, the physical and psychological safety of subjects should be the priority. For neural control devices, deep brain stimulation involves implanted neural electrodes and is associated with a risk of incidental neural damage and infection \cite{bronstein2011deep}. Psychiatric side effects include depression, delusion, euphoria, and disinhibition \cite{pinsker2013psychiatric}. Transcranial magnetic stimulation, on the other hand, is noninvasive and carries an exceptionally rare risk of seizure \cite{been2007use}. As these and other technologies are refined and created, improved safeguards against tissue damage and adverse psychological effects must remain a high priority. In current practice, none of these harms is intended either as a goal or as a means; researchers are therefore not directly maleficent, but may nevertheless be indirectly responsible for their subjects' suffering. As the potential for mind control is explored, therefore, efforts should be made to reduce the potential for indirect maleficence and safeguard against the possibility of direct maleficence.

\paragraph{Beneficence.}
In addition to harm prevention, the principle of human welfare should guide mind control efforts. In applied brain stimulation, clinical symptoms can be reduced and improvements in cognitive performance in healthy individuals can be produced \cite{hamilton2011rethinking}. The general trade-offs of mind control strategies are as yet unknown, but several specific trade-offs are already known to exist \cite{frank2007hold,krause2013effect,sarkar2014cognitive,iuculano2013mental}. Concerning control engineering, maximizing benefits and protecting against harm may be at odds. Multi-cost control optimization is harder than single-cost optimization with constraints \cite{bertsekas1995dynamic}, so one approach to minimizing undesirable side-effects is to optimize one cognitive goal while constraining other cognitive functions to remain outside an undesirable range. This effort may also improve our understanding of the dynamic trade-offs between cognitive systems more generally. As theoretical models of brain function at multiple scales develop, new opportunities for maximally beneficent control should be conceptualized and tested. Nonetheless, given early misuses of lobotomies and electroconvulsive therapy, it is important to separate uses of mind control to improve mental function in clinical cases from its potential misuses for social normalization of incarcerated or otherwise vulnerable persons.

\paragraph{Justice.}
The principle of justice demands that the applications of mind control neither participate in nor exacerbate systems of inequality or exploitation. As such, already existing financial barriers to neural control based treatments and enhancements should be reduced. In turn, access to information about the benefits and risks of such applications should be judiciously increased, in concert with efforts to improve general public health education, especially in underserved communities and countries. To date, mind control application exists in relatively circumscribed clinical and experimental contexts. If, however, mind control strategies for optimal cognitive function become widely available, or even marketable for public consumption, safeguards should preclude the construction of an \emph{enhanced} class, which may result in harm to the \emph{unenhanced} in competitive environments \cite{hamilton2011rethinking}.

\paragraph{Autonomy.}
In the process of guiding the mind, it is imperative that the individual's autonomy, or power of self-determination, remains intact. Autonomy is violated by any involuntary use of mind control on anyone, whether by explicit or implicit coercion. Explicit coercion means forcing individuals to undergo mind control against their conscious will, whether for the perceived greater good of society or the advancement of some constituency. The strongest safeguard against explicit coercion to date has been informed consent \cite{wear1994informed}; however, the informed consent process still needs to be improved \cite{lentz2016paving} and additional methods to supplement it should be developed. This is especially relevant to potential applications of mind control in vulnerable populations. In the future, moreover, should a group of enhanced individuals gain dominance, unenhanced individuals could conceivably be at risk of involuntary medicalization \cite{hamilton2011rethinking}. 

Implicit coercion means manipulating individuals through the activation of external or internal pressure rather than force. While implicit coercion has sullied the history of research ethics \cite{shamoo2009responsible}, the distinction between implicit coercion and participatory incentive today still needs further clarification. For mind control, implicit coercion may include social pressures to increase productivity or efficiency in competitive environments \cite{mccabe2005non,sahakian2007professor,fox2011brain,snowball2013long}. What is more, the reality of structural constraints and mediated freedoms makes appropriate levels of adult autonomy difficulty to calibrate. Nevertheless, research clinicians should make an effort not to capitalize on social conditions that might introduce implicit coercion. For example, poverty may predispose subjects to enter paid research trials, leading to higher rates of experimentation on lower class bodies. As such, the potential for subjects to enroll in research and clinical designs under conditions of implicit coercion should be the focus of redoubled ethical review.

While our discussion here bears on humans with a typical adult capacity for self-determination, further ethical considerations would come into play were mind control to become available for people lacking such capacity -- for example, children, some elderly individuals, and adults with certain forms of mental or physical disabilities.

\section*{Rethinking human persons}
As mind control develops, the ability to interact intelligently with human nature may bring certain stakes into sharper focus. Humans privilege the notion of a ``mind'' and perceived internal locus of control as central to their identities \cite{wilson2015extended}. Further, within minds, humans privilege some traits, such as social comfort, honesty, kindness, empathy, and fairness, as more fundamental than functions, such as concentration, wakefulness, and memory \cite{riis2008preferences}. These different values depend on the notion of conscious identity and are often at the core of common ethical distinctions applied to humans \emph{versus} other animals \cite{olson1999human}. Importantly, modern notions of human persons, influenced by continuing advances in the cognitive and brain sciences, erode the classical boundary between the ethical treatment of humans and animals \cite{singer2011practical}. These theories suggest that tolerance for mind control may scale with ethical issues identified across all animals. In kind, this implies that there exists a need to evaluate the practice of mind control on both human and non-human animals and its permissible scope in principle. 

As the science of mind control advances, it will be important to clarify acceptable control practices with respect to our fundamental nature and self-identity. In addition, the potential for mind control to undermine responsibility connects to our fundamental intuitions about whether we really control what we do \cite{dennett2015elbow}. Implicit within this new technology, then, is the call to redefine ourselves. For this reason, scientists, clinicians, ethicists, and philosophers will need to work together.

\section*{Conclusions}
The study of mind control in human brains has been developing over several years without any particular name. Here, we have described mind control to be fundamentally a problem of network control in the human brain. New frontiers at the intersection of network neuroscience and cognitive science can provide striking new questions and possibilities in understanding and controlling cognitive function. Our success will be predicated on the development of robust theories of neural coding and advances in technologies for recording and manipulating neural activity. In reclaiming a term long relegated to science fiction, new opportunities within and between computational, cognitive, clinical, ethical, and control neuroscience may produce a new era in the science of guiding the mind.

\newpage
\section*{Acknowledgments}
The authors thank Joshua Gold, Elisabeth Karuza, Richard Betzel, and the anonymous reviewers for helpful comments and discussion regarding this work. JDM acknowledges support from the Office of the Director at the National Institutes of Health through grant number 1-DP5-OD-021352-01. DSB acknowledges support from the John D. and Catherine T. MacArthur Foundation, the Alfred P. Sloan Foundation, the Army Research Laboratory and the Army Research Office through contract numbers W911NF-10-2-0022 and W911NF-14-1-0679, the National Institute of Health (2-R01-DC-009209-11, 1R01HD086888-01, R01-MH107235, R01-MH107703, R01MH109520, 1R01NS099348 and R21-M MH-106799), the Office of Naval Research, and the National Science Foundation (BCS-1441502, CAREER PHY-1554488, BCS-1631550, and CNS-1626008). WS-A acknowledges support from  the John Templeton Foundation. The content is solely the responsibility of the authors and does not necessarily represent the official views of any of the funding agencies. 
\newpage
		
\bibliographystyle{naturemag}
\bibliography{JDMReferences}
		
\end{article}

\end{document}